\documentclass[showpacs,showkeys,amsmath,amssymb,superscriptaddress,reprint, preprintnumbers,nofootinbib,notitlepage, groupdaddress,showabstract,aps,prl,twocolumn]{revtex4-1}

\pdfoutput=1 
             
\usepackage[font=small]{caption}
\usepackage[utf8]{inputenc}
\usepackage[english]{babel}
\usepackage{bm}
\usepackage{graphicx}
\usepackage{dcolumn}
\usepackage{pbox}
\usepackage{epsfig}
\usepackage[dvipsnames]{xcolor}
\usepackage{slashed}
\usepackage{amssymb}
\usepackage{mathrsfs}
\usepackage{color}
\usepackage[font=small]{caption}
\usepackage[font=small]{subcaption}
\usepackage{url}
\definecolor{DarkBlue}{rgb}{0.7, 0.4, 1} 
\definecolor{Blue}{rgb}{0, 0.8, 0} 
\definecolor{MyLightBlue}{rgb}{0.5,0.7,1.9}
\definecolor{MyGreen}{rgb}{0.0,0.2, 0.0}
\definecolor{MyBrickRed}{rgb}{0, 0.5, 0.2}
\RequirePackage{hyperref}
\hypersetup{colorlinks, citecolor=Blue,linkcolor=DarkBlue, urlcolor=Green}
\usepackage[normalem]{ulem}

\newcommand{\bea}{\begin{eqnarray}}
\newcommand{\eea}{\end{eqnarray}}

\makeatletter

\renewcommand\@makecaption[2]{%
  \par
  \vskip\abovecaptionskip
  \begingroup
  
   \small\rmfamily
    \begingroup
     \samepage
     \flushing
     \let\footnote\@footnotemark@gobble
     \@make@capt@title{#1}{#2}\par
    \endgroup
  \endgroup
  \vskip\belowcaptionskip
}
\makeatother

\DeclareUnicodeCharacter{2212}{-}
\begin{document}
\title{Lepton number violation from Higgs/$Z$ decays into a pair of right-handed neutrinos } 
\author{Arindam Das}
\email{adas@particle.sci.hokudai.ac.jp}
\affiliation{Institute for the Advancement of Higher Education, Hokkaido University, Sapporo 060-0817, Japan}
\affiliation{Department of Physics, Hokkaido University, Sapporo 060-0810, Japan} 
\author{Takaaki Nomura}
\email{nomura@scu.edu.cn}
\affiliation{College of Physics, Sichuan University, Chengdu 610065, China} 
\author{Kei Yagyu}
\email{yagyu@rs.tus.ac.jp}
\affiliation{Department of Physics, Tokyo University of Science, 1-3,Kagurazaka, Shinjuku-ku, Tokyo 162-8601, Japan}
\begin{abstract}  
We explore signatures of Lepton Number Violation (LNV) from decays of the discovered Higgs ($h$) and Z bosons into a pair of Right-Handed Neutrinos (RHNs). 
Due to the Majorana nature of RHNs, the final state can be the same-sign dilepton plus jets leading to 2 units of LNV. 
As a simple but plausible scenario, we investigate such a signal in models with a new $U(1)_X$ gauge symmetry which naturally introduces three RHNs for gauge anomaly cancellation, and is spontaneously broken down by a vacuum expectation value of an isospin singlet scalar field ($\phi$). 
In this scenario, $h$ and the Z boson can decay into a pair of RHNs via the $h$-$\phi$ mixing and the $Z$-$Z'$ mixing with $Z'$ being a new massive gauge boson, respectively. 
Estimating the LNV signal and corresponding backgrounds for the $\ell^{\pm} \ell^{\pm} 4j$ 
final states, we find bounds on the $h$-$\phi$ mixing and the $Z$-$Z^\prime$ mixing as a function of a mass of RHNs at Higgs factories, e.g., at High-Luminosity LHC with $\sqrt{s}=$ 14 TeV, future $e^-e^+$ colliders at $\sqrt{s}=$ 250 GeV, muon colliders at $\sqrt{s}=$ 125 GeV as well as at Z factories, e.g., CEPC and FCC-ee at $\sqrt{s}=$ 91.2 GeV. We also discuss limits on the scalar mixing at 
$e^-\mu^+$ and $\mu^+ \mu^+$ collisions ($\mu$TRISTAN) with $\sqrt{s}=$346 GeV/775 GeV and 2 TeV, respectively.

\end{abstract}
\maketitle
\noindent

\noindent
{\textbf{Introduction}--} The long standing puzzle of the origin of tiny neutrino masses observed by various experiments~\cite{ParticleDataGroup:2024cfk} allows us to step forward beyond the premise of Standard Model (SM) scenarios. Among many interesting aspects, models with a general $U(1)_X$ gauge symmetry~\cite{Oda:2015gna,Das:2016zue,Das:2017flq} provide one of the simplest realizations for generating tiny neuitrino masses through the seesaw mechanism~\cite{Minkowski:1977sc,Yanagida:1979as,Gell-Mann:1979vob,Mohapatra:1979ia,Schechter:1980gr}, because three Right-Handed Neutrinos (RHNs) are inevitably introduced to cancel gauge anomalies. We can alternatively describe the $U(1)_X$ symmetry as a linear combination of anomaly free $U(1)_Y$ and $U(1)_{\rm B-L}$ \cite{Davidson:1978pm,Marshak:1979fm} gauge groups. 
In this scenario, the Higgs sector is extended to have an additional SM-singlet scalar field ($\phi$) charged under $U(1)_X$ whose Vacuum Expectation Value (VEV) gives masses of RHNs as well as a new massive gauge boson $Z'$. 

Despite their significance, it is quite difficult to directly discover RHNs at collider experiments in the minimal type-I seesaw model without the $U(1)_X$ symmetry. 
This is mainly due to their very large masses relative to the electroweak scale, or their extremely suppressed couplings with the SM sector for lighter RHNs.
In contrast, in our $U(1)_X$ scenario, 
RHNs can be easily produced via the decays of the discovered Higgs boson $(h)$ and the $Z$ boson, where the former (latter) is realized due to a scalar mixing between $h$ and $\phi$ (a gauge mixing between $Z$ and $Z'$). 
In addition, Lepton Number Violation (LNV), another important ingredient in proving the seesaw mechanism, can also be tested via the decay of RHNs into same-sign dileptons plus multi-jets, which is allowed by the Majorana nature of RHNs. 
Therefore,  models with $U(1)_X$ provide not only the simple and natural scenario to realize the seesaw mechanism but also its testable framework at collider experiments.   
The lepton number violating decays of the Higgs boson have been discussed in the left-right symmetric models~\cite{Maiezza:2015lza}, and the minimal type-I seesaw model with a singlet scalar field~\cite{Yang:2025jxc} at Muon Colliders (MCs). 

In this Letter, we discuss the signature of LNV via the decays of $h$ and $Z$ bosons into a pair of RHNs at Higgs factories and Z factories in the $U(1)_X$ model. 
Production cross section of $h$ at LHC with a center of mass energy $\sqrt{s}$ of $\sqrt{s}=13$ (14) TeV 
has been computed to be 48.61 (54.72) pb \cite{Anastasiou:2016cez,Cepeda:2019klc} at N$^3$LO QCD and NLO EW from the gluon fusion process. 
At future electron-positron colliders with $\sqrt{s}=250$ GeV \cite{FCC:2018byv,
deBlas:2019rxi,EuropeanStrategyGroup:2020pow,FCC:2025lpp}, the Higgs production in association with $Z$ boson could reach at around 300~\cite{Asner:2013psa,ILC:2013jhg} (230~\cite{Sun:2016bel}) fb for polarized (unpolarized) electron-positron beams with polarization $P(e^-, e^+)=(-0.8,0.3)$. In addition, MCs at $\sqrt{s}=125$ GeV could produce $h$ in the $s$-channel process~\cite{Barger:1995hr,Han:2012rb,Greco:2016izi,deBlas:2022aow} with a cross section of 20 pb \cite{deBlas:2022aow} taking Initial State Radiation (ISR) and Beam Energy Spread (BES) into consideration.
On the other hand, $Z$ factories~\cite{Antusch:2025lpm} can be realized at electron-positron colliders 
with $\sqrt{s}= 91.2$ GeV, by which the cross section is significantly enhanced by the on-shell resonance of the $Z$ boson, and  $\mathcal{O}(10^{12})$ number of $Z$ bosons can be produced \cite{ALEPH:2005ab,CEPC-SPPCStudyGroup:2015csa,CEPCStudyGroup:2018ghi,CEPCStudyGroup:2023quu,Ai:2025cpj,CEPCStudyGroup:2025kmw}.
Recently, $e^-\mu^+$ and $\mu^+\mu^+$ colliders have also been considered as alternative possibilities of the Higgs factory, the so-called $\mu$TRISTAN~\cite{Hamada:2022mua}. 
The Higgs boson can be produced via the weak boson fusion process, and their cross sections are given to be about 91 (4) fb from the $W$ ($Z$) boson fusion at $e^-\mu^+$ collisions with $\sqrt{s} = 346$ GeV and 1 ab$^{-1}$ of the integrated luminosity (${\cal L}$), while 
the cross section can be 54 fb from the $Z$ boson fusion at $\mu^+\mu^+$ collisions with $\sqrt{s} = 2$ TeV and ${\cal L}=$100 fb$^{-1}$. 
In addition, $e^- \mu^+$ collisions with $\sqrt{s}= 775$ GeV and ${\cal L}=$1 ab$^{-1}$ have also been proposed as $\mu$Tevatron in~\cite{Hamada:2022mua}, where the Higgs production cross section is 472 (20) fb from the $W$ $(Z)$ boson fusion.

We study the final states with LNV from the Higgs and Z decays into $\ell^\pm \ell^\pm 4j$ at the Higgs and Z factories, and simulate the SM Backgrounds (BGs). 
We then estimate bounds on the scalar mixing and $Z$-$Z^\prime$ mixing with respect to the mass of RHNs at Higgs and $Z$ factories.

\noindent
{\textbf{Framework}--}
Under the $SU(2)_L\otimes U(1)_Y \otimes U(1)_X$ gauge symmetry, the SM quarks, leptons and Higgs fields are charged as follows
\begin{align}
&q_L^i\sim \left(2,\frac{1}{6}, \frac{x_H}{6} + \frac{x_\Phi}{3} \right),~ 
u_R^i\sim \left(1,\frac{2}{3},\frac{2}{3} x_H +\frac{x_\Phi}{3}\right),\notag\\
&d_R^i\sim \left(1,-\frac{1}{3}, -\frac{x_H}{3} +\frac{x_\Phi}{3}\right),~
\ell_L^i \sim \left(2,-\frac{1}{2}, -\frac{x_H}{2}-x_\Phi\right), \notag\\
&e_R^i\sim \left(1,-1, -x_H-x_\Phi\right),~
H\sim \left(2,\frac{1}{2}, \frac{x_H}{2}\right),  
\end{align}
where the index $i$ denotes the flavor $i=1, 2, 3$. 
As new fields, we introduce three generations of RHNs $N_R^i$ to cancel gauge and mixed gauge-gravity anomalies and an isospin singlet scalar $\Phi$ being transformed as 
\begin{align}
N_R^i=(1,1,0,-x_\Phi),~~\Phi=(1,1,0, 2 x_\Phi). 
\end{align}
Generally, the kinetic-mixing term between $U(1)_Y$ and $U(1)_X$ appears, 
but we simply neglect this term due to its smallness.

The scalar potential is given by
\begin{align}
  V  &= -\ m_H^2(H^\dag H)  - m_\Phi^2 (\Phi^\dag \Phi) \notag\\
&  + \lambda_H^{} (H^\dag H)^2
  + \lambda_\Phi^{} (\Phi^\dag \Phi)^2 
  +\lambda_{\rm mix} (H^\dag H)(\Phi^\dag \Phi)~.
\end{align}
In the unitary gauge, $H$ and $\Phi$ are parameterized as
\begin{align}
  H\ = \ \frac{1}{\sqrt{2}}\begin{pmatrix} 
  0 \\
  v + \tilde{h}
  \end{pmatrix}~, \quad {\rm and}\quad 
  \Phi \ =\  \frac{v_\Phi^{} + \tilde{\phi} }{\sqrt{2}}~,\label{eq:scalars}
\end{align}
where $v$ and $v_\Phi^{}$ are the VEVs with $v \simeq 246$ GeV. 
We consider $v_\Phi\gg v$ following LEP bounds discussed in \cite{Das:2021esm}.
After imposing the stationary condition of the scalar potential, we obtain the mass-squared matrix for the scalar fields:  
\begin{align}
M^2=\begin{bmatrix}
2\lambda_H v^2 &  \lambda_{\rm mix} v  v_\Phi \\
\lambda_{\rm mix} v v_\Phi  &  2\lambda_{\Phi}v_\Phi^2 \\
\end{bmatrix},
\end{align}
in the basis of $(\tilde{h}, \tilde{\phi})$ defined in Eq.~(\ref{eq:scalars}). 
These states are related to the mass eigenstates ($h,\phi$) via the $2\times2$ orthogonal rotation matrix $O_R$ as
\begin{align}
\begin{bmatrix}
\tilde{h} \\
\tilde{\phi} \\
\end{bmatrix}=O_R
\begin{bmatrix}
h \\
\phi 
\end{bmatrix}=
\begin{bmatrix} \cos\alpha  &  -\sin\alpha \\
\sin\alpha  &  \cos\alpha  \\  \end{bmatrix}
\begin{bmatrix}
h \\
\phi \\
\end{bmatrix},
\label{eq:rotation}
\end{align}
where $\alpha$ is the mixing angle between these two scalars. The rotation matrix satisfies $O_R^T M^2 O_R = \text{diag}\left(m_{h}^2,m_{\phi}^2\right)$. Using these relations, we solve for the parameters $\lambda_H$, $\lambda_\Phi$ and $\lambda_{\rm mix}$ in terms of $\alpha$ and the scalar masses as
\begin{align}
 \lambda_{H} &= \frac{1}{2v^2}\left(m_{h}^2\cos^2\alpha+m_{\phi}^{2}\sin^2\alpha\right), \nonumber \\
 \lambda_{\Phi}& = \frac{1}{2v_\Phi^2}\left(m_{h}^2\sin^2\alpha+m_{\phi}^{2}\cos^2\alpha\right) , \nonumber \\
 \lambda_{\rm mix}&=\left(\frac{m_{h}^{2}-m_{\phi}^{2}}{2v_\Phi v}\right)\sin2\alpha, 
 \label{coup}
\end{align}
where we identify $h$ with the discovered Higgs boson with $m_{h}\simeq 125$ GeV, and  
$\phi$ with the BSM Higgs boson with its mass being a free parameter. 

The kinetic terms for the scalar fields are given by 
\begin{align}
\mathcal{L}_{\rm kin} = |\mathcal{D}_\mu H|^2 + |{\mathcal{D}_\mu}\Phi|^2. 
\end{align}
The mass matrix for the massive neutral gauge bosons is extracted as  
\bea
M_{ZZ^\prime}= 
\begin{pmatrix}
M_Z^2 & -\frac{v^2}{2}g_Z^{} g_X^{} x_H^{}\\
-\frac{v^2}{2}g_Z^{} g_X^{} x_H^{}  & M_{Z'}^2
\end{pmatrix}, 
\eea
where 
\begin{align}
    M_Z^2 &= \frac{g_Z^2}{4} v^2, \\
    M_{Z'}^2 &= \frac{g_X^2}{4} (x_H^2 v^2 + 16 x_\Phi^2 v_\Phi^2), 
\end{align}
with $g_X^{}$ being the $U(1)_X$ gauge coupling and $g_Z^{} = g/\cos\theta_W$ and $\theta_W$ being the weak mixing angle. 
For $M_{Z'} \gg M_Z$, the masses of $Z$ and $Z'$ are approximately given by the (1,1) and (2,2) elements of the above matrix, respectively, and their mixing angle $\zeta$ is given by 
\begin{align}
    \zeta \simeq 2 \frac{M_Z^2}{M_{Z^\prime}^2}\frac{g_X^{} x_H^{}}{g_Z}. 
\end{align}
For instance, $\zeta$ is given to be of order $10^{-4}$ for $M_{Z'} = 10$ TeV and $g_X^{}x_H \simeq g_Z$.  

\begin{figure*}[!t]
\includegraphics[width=8.6cm]
{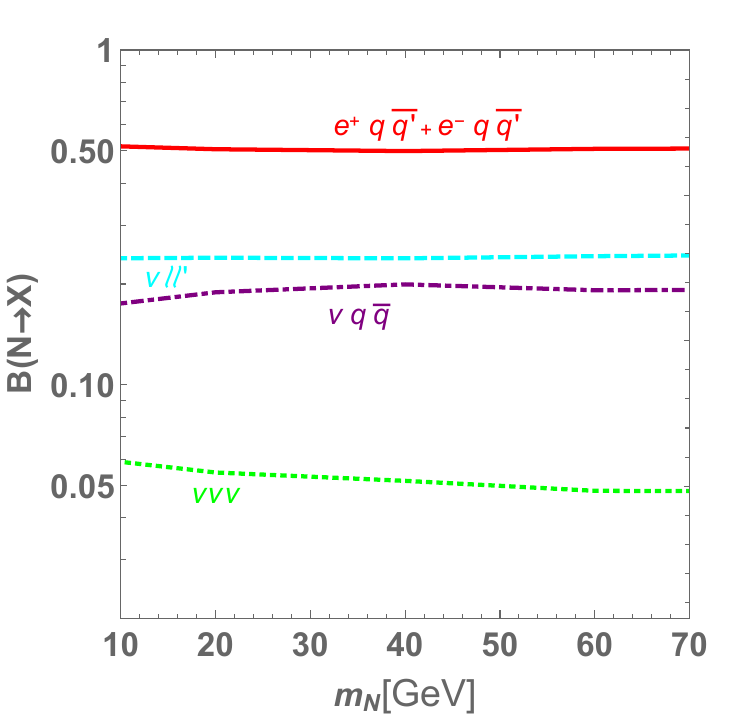}  
\includegraphics[width=8.5cm]{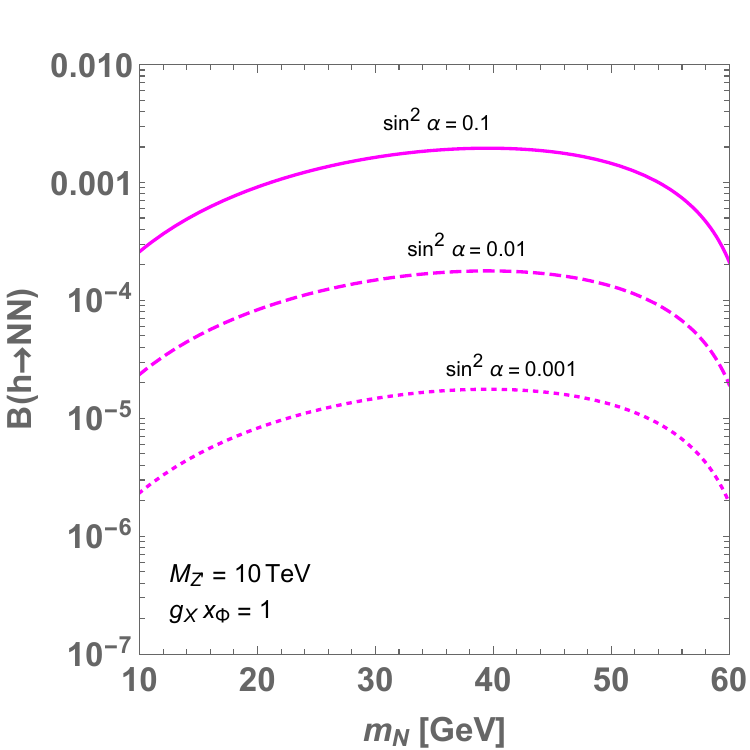}
\caption{Left: Branching ratios of $N$ as a function of $m_N$, where $q,q'\neq t$. Right: Branching ratios of $h\to NN$ as a function of $m_N$ for $\sin^2\alpha = 0.1$ (solid), 0.01 (dashed) and 0.001 (dotted).  }
\label{fig:Ndecay}
\end{figure*}

The relevant Yukawa interactions to the neutrino mass generation can be written as,
\bea
{\cal L} &\supset& - Y_{\nu} \overline{\ell_L} \tilde{H}\,N_R- \frac{1}{2}Y_{N} \Phi \overline{(N_R)^c} N_R + {\rm H.c.,} 
\label{eq:LYk}   
\eea
where $Y_N$ and $Y_\nu$ are $3\times 3$ Yukawa matrices with the former being taken to be a diagonal form without loss of generality, and $\tilde{H} = i \tau^2 H^*$ with $\tau^2$ being the second Pauli matrix. 
From Eq.~\eqref{eq:LYk}, the Majorana mass for the RHNs are generated by the spontaneous breaking of $U(1)_X$ symmetry whereas the Dirac mass for the left-handed neutrinos are generated after the electroweak symmetry breaking as
\begin{equation}
    m_{N} = \frac{Y_{N}}{\sqrt{2}} v_\Phi, \, \, \, \, \,
    m_{D} = \frac{Y_{\nu}}{\sqrt{2}} v\,.
\label{eq:mDI}
\end{equation}
which induces the seesaw mechanism to generate the light active neutrino masses as $-m_D^{} m_{N}^{-1} m_D^T$~\cite{Gell-Mann:1979vob, Sawada:1979dis, Mohapatra:1980yp}. It could successfully explain the origin of tiny neutrino masses and their flavor mixing.

The RHNs do not interact with the SM sector directly, but from the seesaw mechanism, we can write the light neutrino flavor eigenstates $(\nu_\ell)$ in terms of the mass eigenstates of the light $(\nu_m)$ and heavy $(N_m)$ neutrinos as 
\begin{align}
    \nu_\ell \simeq  \nu_m  + V_{\ell N} N_m, 
\end{align}
where $V_{\ell N}(=m_D/m_N)$ is the mixing matrix between the light and heavy mass eigenstates. 
Hence, the Charged Current (CC) and the Neutral Current (NC) interactions in terms of the neutrino mass eigenstates can be written as  
\begin{align}
\mathcal{L}_{\rm CC}  &\supset
 -\frac{g}{\sqrt{2}} W_{\mu}
  \bar{\ell} \gamma^{\mu} P_L V_{\ell N} N_m  + {\rm H.c.}, 
\label{CC} \\
\mathcal{L}_{\rm NC} &\supset  
 -\frac{g_Z^{}}{2}  Z_{\mu} 
\Big[ 
  \overline{N_m} \gamma^{\mu} P_L  (V_{\ell N}^\dagger V_{\ell N'}) N_m' \nonumber \\
&+ \overline{\nu_m} \gamma^{\mu} P_L V_{\ell N}  N_m 
  + {\rm H.c.} \Big]. 
\label{NC}
\end{align} 
where $\ell$ denotes the three generations of the charged leptons in the vector form, and 
$P_L =\frac{1}{2} (1- \gamma_5)$. 
From these interactions, the RHNs can decay into leptonic and semileptonic modes \cite{Das:2022rbl,A:2025ygb}. From Eq.~(\ref{NC}), we see that $Z\to N_m N_m$ mode is suppressed by $|V_{\ell N}|^4$ making it challenging to study at colliders \cite{Das:2017pvt} whereas other interactions including those from Eq.~(\ref{CC}) are proportional to $|V_{\ell N}|^2$. 
In the following discussion, we simply denote the light and heavy neutrinos as $\nu$ and $N$, respectively, by suppressing the subscript $m$.


In the left panel of Fig.~\ref{fig:Ndecay}, the branching ratios of $N$ are shown as a function of $m_N$. 
In this plot, we consider the decay of the lightest heavy neutrino $N$, and $N$ to be assumed being dominantly coupled with the first generation of the SM lepton.
We see that the $N \to \ell^{\pm}q \bar{q}^\prime$ mode is dominant throughout the mass range under consideration.
The current upper limit of $|V_{\ell N}|$ has been found to be $|V_{\ell N}|^2 \simeq (m_D/m_N)^2 \lesssim 5\times 10^{-6}$ from the recent searches~\cite{Das:2023tna,CMS:2024xdq,CMS:2024bni,Das:2024kyk}
for $m_N = \mathcal{O}(10)$ GeV considering prompt decays of heavy neutrinos at the LHC. We adopt this condition in our analysis. Heavy neutrinos could be potentially long-lived if $|V_{\ell N}|^2\simeq (m_D/m_N)^2 < 10^{-6}$ that we do not consider in this Letter.

\noindent
{\textbf{Heavy neutrino production from $h$ and $Z$ decays}--}
In the following discussion, we consider the scenario with $m_{\phi},~M_{Z'} \gg m_h > m_N/2$. 
In this case, $N$ can be produced via the decays of $h$ and $Z$. 

The decay rate of $h \to NN$ is given as 
\begin{align}  
\Gamma(h \to NN)  
= \frac{m_N^2 m_h \sin^2\alpha}{16 \pi v_\Phi^2}  \left(1-4\frac{m_{N}^2}{m_{h}^2}\right)^{3/2}, 
\label{hD}
\end{align}
which will be interesting to study at the Higgs factory considering $\sin^2\alpha \simeq 0.1$ for $m_{\phi} \gg m_{h}$ \cite{CLIC:2018fvx,Das:2022oyx}. 
Other decay rates of $h$ into SM particles are given by the corresponding SM value times $\cos^2\alpha$~\cite{Nomura:2024pwr,Li:2025luf}. 
The Higgs boson can also decay into $N \nu$ via the Yukawa coupling $Y_\nu$, and the decay rate is given by  
\bea
\Gamma(h\to N \bar{\nu})= \frac{|V_{\ell N}|^2 m_N^2 m_h}{8\pi v^2} \Big(1-\frac{m_N^2}{m_h^2}\Big)^2.
\eea
Hence, the branching ratio of $h \to NN$ is calculated as  
\begin{align}
    &\mathcal{B}(h\to N N) \notag\\
&= 
\frac{\Gamma(h\to NN)}{\Gamma^{\rm{SM}}_{h}\cos^2\alpha +\Gamma(h\to NN)+2\Gamma(h\to N\nu)},
\label{BR}
\end{align}
where $\Gamma^{\rm{SM}}_{h}=4.1$ MeV is the total width of the SM Higgs boson~\cite{LHCHiggsCrossSectionWorkingGroup:2016ypw}, and the factor 2 comes in the denominator combining $h\to N\nu$ and $h\to N\bar{\nu}$ modes \cite{
BhupalDev:2012zg,Das:2017zjc,Das:2017rsu} considering $|V_{\ell N}|^2\simeq 5\times 10^{-6}$. 
We show the branching ratio of $h \to NN$ as a function of $m_N$ for different $\sin^2\alpha$ in the right panel of Fig.~\ref{fig:Ndecay}. We find that $\mathcal{B}(h\to NN)$ can be maximum for $m_N\simeq 40$ GeV due to its dependence on $m_N$ as shown in Eq.~(\ref{hD}).
We then obtain the lepton number violating signal from the Higgs decay chain of $h \to N N \to \ell^\pm \ell^\pm 4 j$.

Next, let us discuss the decay of $Z \to NN$. As aforementioned, 
the $\bar{N}NZ_\mu$ coupling from the NC is highly suppressed by $|V_{\ell N}|^2$, so that its contribution to the decay is negligibly small. However, there is another source giving the $\bar{N}NZ_\mu$ coupling via the $Z$-$Z'$ mixing. Namely, 
\bea
\mathcal{L}_{\rm int} \supset 
\frac{x_\Phi g_X}{2} \bar{N} \gamma^\mu \gamma^5 N ( Z_\mu\sin\zeta + Z'_\mu \cos\zeta).
\eea
Hence, the decay rate of $Z \to NN$ is given by 
\bea
\Gamma(Z\to NN) \simeq \frac{(x_\Phi g_X)^2 \zeta^2}{24 \pi} M_Z \left(1-4 \frac{m_N^2}{M_{Z}^2}\right)^{3/2}. 
\label{RHNdecay}
\eea
Therefore, the branching ratio of $B \to NN$ is calculated as 
\bea
\mathcal{B}(Z\to N N)= 
\frac{\Gamma(Z\to NN)}{\Gamma_Z^{\rm SM} + \Gamma(Z\to NN)},  
\eea
with $\Gamma_Z^{\rm SM} \simeq 2.5$ GeV being the total width of the Z boson in the SM~\cite{Beenakker:1988pv,ALEPH:2006bhb,ALEPH:2010aa}. 
The magnitude of the $Z$-$Z'$ mixing is constrained to be $|\zeta|< {\cal O}(10^{-3})$ from the electroweak precise measurements~\cite{Erler:2009jh}.
Putting $x_\Phi g_X^{} = 1$ and $\zeta = 10^{-3}$, 
the branching ratio of $Z \to NN$ is estimated to be about $5\times 10^{-7}$. 
This could be tested at $Z$ factories with ${\cal O}(10^{12})$ of the production of the $Z$ boson. 
Similar to the Higgs decay, we can prove the LNV via the decay chain of $Z \to NN \to \ell^\pm \ell^\pm 4j$.

\begin{figure*}[htb]
\begin{center}
\includegraphics[width=9cm]{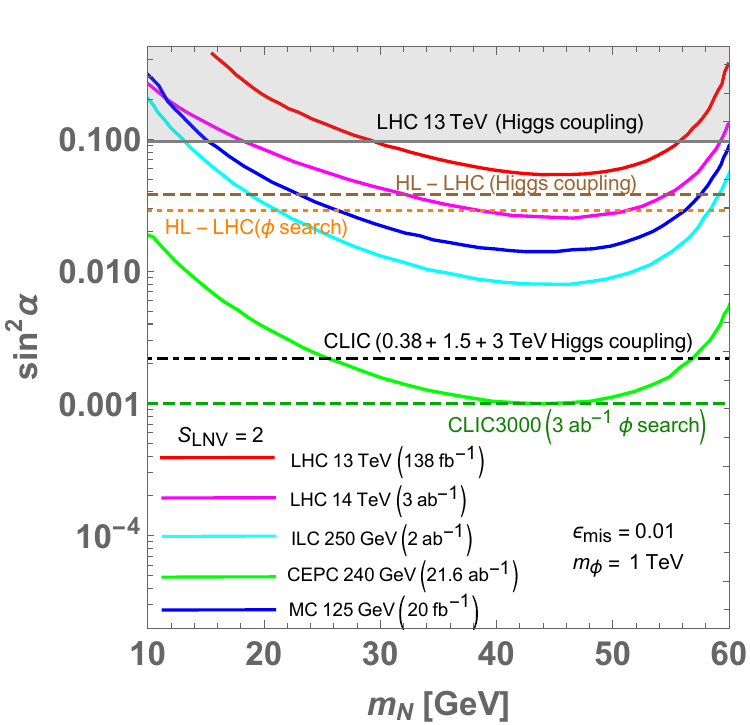}  
\includegraphics[width=8.6cm]{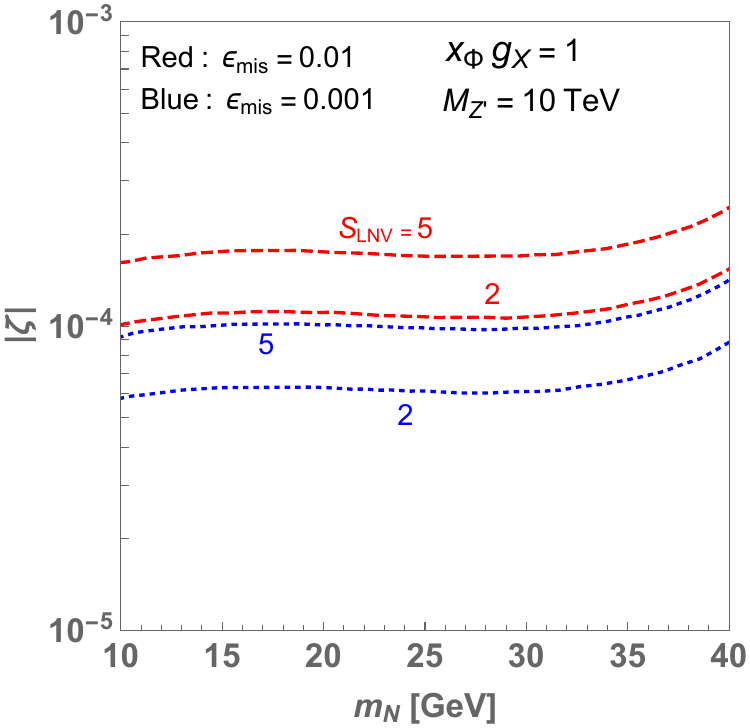}  
\caption{Left: LNV signature can be explored from the Higgs decay with the significance $S_{\rm LNV} = 2$ on each curve
in the $m_N^{}$-$\sin^2\alpha$ plane. 
The horizontal lines show the upper limit on $\sin^2\alpha$ from the Higgs coupling measurements and direc searches for $\phi$. 
Right: LNV signature can be explored from the $Z$ decay with the significance $S_{\rm LNV} = 2$ and 5 assuming $\epsilon_{\rm mis}=0.01$ (red) and $0.001$ (blue)
on each curve in the $m_N^{}$-$|\zeta|$ plane. }
\label{fig:Sig}
\end{center}
\end{figure*}

\noindent
{\textbf{Results and discussion}--}
\begin{table}[t]
\begin{tabular}{c||c c c}\hline  
&  LHC& ~$e^+e^-$ colliders & $\mu^+\mu^-$ colliders  \\\hline
$\sqrt{s}$&  13 (14) TeV& ~250 (240) GeV& 125 GeV   \\
${\cal L}$ [ab$^{-1}$]&  0.138 (3) & ~2 (21.6)& ~0.02   \\
$\sigma_h^{\rm SM}$ [pb] & 48.61 (54.72) & 0.23 (0.20) & 20 \\
BG process   & $\ell^+\ell^-4j$ & $\ell^+\ell^-jjZZ(ZW)$ & $\ell^+\ell^-4j$  \\
$\sigma_{\rm B}$ [fb]  & $365~(447)$  & ${\cal O}(10^{-5})$ & $10$  \\
$\epsilon_{\rm eff}$  & ${\cal O}(0.1)$ & $0.2$-0.6 & $0.2$-0.6  \\\hline
\end{tabular}
\caption{Simulation for the LNV signature at each Higgs factory. For $e^+e^-$ colliders, $\sigma_h^{\rm SM}$ indicates $hZ$ production with unploralized beam.}\label{tab:1}
\end{table}
We estimate the signal significance for the LNV signal $h/Z \to NN \to \ell^\pm \ell^\pm 4j$ by performing parton level simulation. 
We first consider the $h$ decay.  
For the Higgs production, we consider $gg \to h$, $e^+e^- \to h Z$ and $\mu^+ \mu^- \to h$ at the LHC, future electron-positron colliders and MCs, respectively. For the $hZ$ process, we consider hadronic decays of the $Z$ boson. 
At MCs, we take into account the effects of ISR and BES~\cite{deBlas:2022aow}. 
The production cross sections are given in Table~\ref{tab:1}. 


In order to simulate signals and BG processes, we use {\tt MadGraph5} \cite{Alwall:2011uj}. 
The significance of the LNV signal is defined by 
\begin{equation}
S_{\rm LNV} = \frac{\epsilon_{\rm eff} N_{\rm S} }{\sqrt{\epsilon_{\rm eff} N_{\rm S} + 2 \epsilon_{\rm miss} N_{\rm B}}}
\label{eq:significance},
\end{equation}
where $N_{\rm S}~(N_{\rm B})$ is the number of signal (BG) events,  
$\epsilon_{\rm eff}$ denotes the efficiency for the signal events and 
$\epsilon_{\rm miss}$ denotes the charge misidetification rate explained below (the factor 2 appears since we count misidentification of both $\ell^+$ and $\ell^-$). 
The number of signal events is estimated by 
\begin{align}
N_{\rm S} &= \sigma_h^{\rm SM}\cos^2\alpha\times \mathcal{B}(h \to NN) \notag\\
&\times[\mathcal{B}^2(N \to \ell^+ jj) + \mathcal{B}^2(N \to \ell^- jj)],
\end{align}
where $\ell = e,\mu$ and $\sigma_{h}^{\rm SM}$ is the $h$ production cross section in the SM, see  Table~\ref{tab:1}.
Regarding the BG, we consider the following processes at each collider: 
\begin{align}
&pp \to \ell^+ \ell^- 4j, \notag\\
&e^+e^- \to \ell^+ \ell^- jjZZ/\ell^+ \ell^- jjWZ, \notag\\
&\mu^+\mu^- \to \ell^+ \ell^- 4j, 
\end{align}
where 
one of the leptons is charge-misidentified with the rate of $\epsilon_{\rm miss}$ at detectors. 
We impose the following kinematical cuts at LHC and MCs;
\begin{align}
&p_T(\ell) > 5~\text{GeV},\quad 
p_T(j) > 15~\text{GeV},\notag\\
& 110~\text{GeV} < M_{\ell^+ \ell^- 4j} < 140~\text{GeV}, 
\end{align}
where $p_T(\ell)$ and $p_T(j)$ are respectively the transverse momentum of a lepton and a jet, and $M_{\ell^+ \ell^- 4j}$ denotes the invariant mass for the $\ell^+ \ell^- 4j$ system.
On the other hand, we impose 
\begin{align}
&p_T(\ell) > 5~\text{GeV},\quad 
p_T(j) > 5~\text{GeV}, \label{eq:cut_z}
\end{align}
at $e^+e^-$ colliders. 
%
%
%
%
In Table~\ref{tab:1}, we summarize the center of mass energy, integrated luminosity (${\cal L}$), 
Higgs production cross section, BG process, 
BG cross section under the cuts and signal efficiency $\epsilon_{\rm eff}$ for each collider experiment.

In the left panel of Fig.~\ref{fig:Sig}, we show the results 
for the LNV signal at the Higgs factories. 
Points on each solid curve satisfy $S_{\rm LNV} =2$ on the $m_N$-$\sin^2\alpha$ plane, where we adopt $\epsilon_{\rm mis} =0.01$ for all colliders.  
We thus find that current LHC data can test $\sin^2 \alpha \simeq 0.06$ for $m_N \simeq 40$ GeV. Furthermore, the sensitivity for $\sin^2 \alpha$ can reach around $ 10^{-2}$  by HL-LHC, ILC and MC, and $ 10^{-3}$ by CEPC (upgraded) for $m_N \simeq 40$ GeV.
Due to the nature of $\mathcal{B}(h\to NN)$ and its dependence on $m_N$ we find that strongest limit on the scalar mixing could be obtained for $m_N\simeq 40$ GeV.
For comparison, we also show current and future perspective for testing $\sin^2\alpha$ from the Higgs coupling measurements at LHC Run-II~\cite{ATLAS:2020qdt}, HL-LHC~\cite{Cepeda:2019klc,CLIC:2018fvx}, CLIC~\cite{CLIC:2018fvx} and from direct searches for $\phi$ assuming $m_{\phi} = 1$ TeV at the HL-LHC~\cite{Cepeda:2019klc,CMS:2018qmt,CLIC:2018fvx} and CLIC~\cite{CLIC:2018fvx,Buttazzo:2018qqp}. 

In addition, we study limits on $\sin^2\alpha$ taking $m_N = 40$ GeV at $e^-\mu^+$ and $\mu^+\mu^+$ colliders 
with $(\sqrt{s},{\cal L})=(346$ GeV, 1 ab$^{-1}$) and (2 TeV, 100~fb$^{-1}$), respectively, under BG free scenarios ($\mu$TRISTAN). 
The Higgs boson can mainly be produced via the $W$ boson fusion process and its cross section is given to be about 91 fb at the $e^-\mu^+$ collider. 
Using Eq.~(\ref{eq:significance}), we obtain the upper limit of $\sin^2\alpha = 0.053$ which will be comparable with the expectation given at the HL-LHC.
In a similar fashion, we obtain the upper limit of $\sin^2\alpha = 0.092$ from the $\mu^+\mu^+$ collider, where the Higgs boson is produced via the $Z$ boson fusion and its cross section is given to be about 54 fb.  
For the $e^-\mu^+$ collision with  $(\sqrt{s},{\cal L})=(775$ GeV, 1 ab$^{-1}$), i.e., $\mu$Tevatron, 
the Higgs production cross section is 472 fb from the $W$ boson fusion. Using Eq.~(\ref{eq:significance}), we obtain the upper limit of $\sin^2\alpha=0.0097$ which could be compared to the bounds obtained from the ILC from Fig.~\ref{fig:Sig}. 

Finally, we discuss the signal significance of the $\ell^{\pm} \ell^{\pm} 4j$ signal from the $Z$ boson decay into a pair of $N$ at Z factories. 
In particular, we focus on Tera-Z experiments, where $10^{12}$ Z bosons are expected to be produced at the Z-pole energy, corresponding to $\sim 100$ ab$^{-1}$ of the integrated luminosity. 
A possible BG process in the SM is $e^+e^- \to \ell^+ \ell^- 4j$ where it mimics the signal if the detector misidentifies the charge of one lepton. We estimate the cross section applying transverse momentum cut, $p_T > 5$ GeV for charged lepton and jets using {\tt MadGraph5} \cite{Alwall:2011uj} at the parton level.
After the kinematical cuts given in Eq.~(\ref{eq:cut_z}), we obtain the BG cross section as
\begin{equation}
\sigma(e^+ e^- \to \ell^+ \ell^- 4j) \simeq 7.2 \ {\rm fb}.
\end{equation}
For the signal events, the efficiency $\epsilon_{\rm eff}$ under the $p_T$ cut is also estimated, and we find $\epsilon_{\rm eff} \simeq 0.1$ to $\epsilon_{\rm eff} \simeq 0.4$ for $m_{N} = 10$ GeV to 40 GeV.
Then, we calculate the signal significance by using Eq.~\eqref{eq:significance}. 
The number of events is estimated by 
\begin{align}
N_{\rm S} &= 4.1 \times 10^{12}\times \mathcal{B}(Z \to NN) \notag\\
&\times[\mathcal{B}(N \to \ell^+ jj)^2 + \mathcal{B}(N \to \ell^- jj)^2],
\end{align}
where $4.1 \times 10^{12}$ is the number of $Z$ boson at CEPC Z-factory~\cite{CEPCStudyGroup:2025kmw}.
In the right panel of Fig.~\ref{fig:Sig}, we show the significance $S_{\rm LNV}$ for the LNV signal at $Z$-factories where red-dashed and blue-dotted curves correspond to charge mis-identification rate $1 \%$ and $0.1 \%$ cases respectively. Thus, the signal can be tested with 90$\%$ confidence level (CL), corresponding to $S_{\rm LNV}=2$, for $|\zeta| \gtrsim 10^{-4}(6\times 10^{-5})$ for $\epsilon_{\rm{mis}} =0.01(0.001)$.

\noindent
{\textbf{Conclusions}--} 
In this work, we have investigated LNV signals from the decays of the 125 GeV Higgs boson $h$ and $Z$ boson into a pair of RHNs ($N$). The decay of $h \to NN$ is induced via the mixing with a singlet scalar which induces Majorana masses for the RHNs after the $U(1)_X$ symmetry breaking. On the other hand, the $Z$ boson decays into $NN$ via the $Z$-$Z'$ mixing. We have performed simulation studies at parton level for the LNV signals and BGs at the Higgs/$Z$ factories. It has been found that LNV signals via the Higgs decay can be explored for the scalar mixing $\sin^2 \alpha \gtrsim  10^{-2} ( 10^{-3})$ at the HL-LHC, MCs and ILC (CEPC) for $m_N \simeq 40$ GeV. 
We also have shown that a similar sensitivity can be obtained at $e^-\mu^+$ and $\mu^+\mu^+$ collisions ($\mu$TRISTAN) as those given at the HL-LHC. 
%
For the LNV signal from $Z$ decay, we can have sensitivity to test it when $Z$-$Z'$ mixing is $|\zeta| \gtrsim 10^{-4}$.
Therefore, we have shown that future Higgs and $Z$ factories are excellent opportunities to find LNV signals from the decays of Higgs/$Z$ providing significant avenues to explore the Majorana nature of neutrinos.

\noindent
{\textbf{Acknowledgements}--} 
We thank Sanjoy Mandal for useful discussions.

{\bf Note added:} For a future work we will be considering  long-lived heavy neutrino pairs produced from Higgs and $Z$ mediated interactions \cite{DNY}.
\vspace{7mm}
\vspace{-0.398in}
\bibliographystyle{utphys}
\bibliography{bibliography}
\end{document}